\documentclass[aip,jcp,amsmath,amssymb,reprint,numeric]{revtex4-1}
\usepackage{graphicx}
\usepackage{subfigure}
\usepackage{color}
\usepackage{amsmath}
\usepackage{bm}

\begin{document}
\title {Determination of onset temperature from the entropy for fragile to strong liquids.}
\author{Atreyee Banerjee}

\affiliation{\textit{Polymer Science and Engineering Division, CSIR-National Chemical Laboratory, Pune-411008, India}}

\author{Manoj Kumar Nandi}
%\thanks{M. K. Nandi and A. Banerjee contributed equally to this work.}
\affiliation{\textit{Polymer Science and Engineering Division, CSIR-National Chemical Laboratory, Pune-411008, India}}
\author{ Srikanth Sastry}
\affiliation{\textit{Theoretical Sciences Unit, Jawaharlal Nehru Centre for Advanced Scientific Research, Jakkur Campus, Bengaluru 560 064, India}}

\author{Sarika Maitra Bhattacharyya}
\email{mb.sarika@ncl.res.in}
\affiliation{\textit{Polymer Science and Engineering Division, CSIR-National Chemical Laboratory, Pune-411008, India}}

\date{\today}

\begin{abstract}
In this paper we establish a connection between the onset temperature of glassy dynamics with the change in the entropy for a wide range of model systems.
 We identify the crossing temperature of 
pair and excess entropies as the onset temperature. Below the onset temperature, the residual
multiparticle entropy(RMPE), the difference between excess and pair entropies, becomes positive. 
The positive entropy can be viewed as equivalent to the larger phase space exploration of the system. 
The new method of onset temperature prediction 
from entropy is less
ambiguous, as it does not depend on any fitting parameter like the existing methods. Our study also reveals the connection between fragility and the degree of
breakdown of the Stokes
Einstein (SE) relation.
\end{abstract}
\maketitle

\section{Introduction}

On fast cooling of liquids, if the crystallization is avoided, they enter the supercooled state. In the supercooled state, many liquids exhibit slow dynamics,
 leading eventually to structural arrest in the form of the glass transition.
 Since the structure of the liquid in the supercooled state does not change much,
 the main question 
arises whether this slow down and eventual arrest/glass transition is kinetically driven \cite{chandler-kcm} or it is a reflection of a thermodynamic event
\cite{RFOT,lubchenko}.

When cooled from high temperature, a glass forming liquid shows a number of characteristic temperatures
over which qualitative changes occur in the dynamics of the system. Around the melting temperature one observes a crossover temperature termed the onset temperature ($T_{onset}$)
where the liquid begins to display non-Arrhenius temperature dependence of the relaxation time \cite{sastry2000onset,richert1998dynamics} and sometimes it can be
 identified from the development of a plateau in the time correlation
functions\cite{kob-2}.
Also, one observes the emergence of spatially heterogeneous dynamics 
near $T_{onset}$\cite{sengupta2013breakdown}.
According to  kinetically constrained models based on the concept of dynamical facilitation, dynamic heterogeneity is an effect of segregation
of fast and slow particles\cite{chandler-kcm,cavagna_review}. 
On the other hand, RFOT theory, based on a thermodynamic perspective, describes dynamic heterogeneity as an effect of entropy driven nucleation.
Indeed, at the onset temperature there are thermodynamic changes and the energy of inherent structures (local energy minima) starts to drop from a near
constant value as the system begins to sample deeper minima in the 
potential energy landscape\cite{srikanth}. 
This temperature is also found to be associated with a change in the nature of the local potential energy minima. \cite{sastry2000onset}
Thus the crossover temperature from `normal' to 'slow' dynamics can be treated as an onset of slow dynamics\cite{tarjus-kivelson,sastry2000onset,srikanth}.
On further cooling, the system reaches a temperature called mode coupling transition temperature $T_c$, at which the power law behavior of the mode coupling
theory (MCT) predicts a divergence of the dynamics \cite{Reichman,unravel}.
Earlier studies have reported that the saddle order nearby extrema of the potential energy surface goes to zero at this temperature 
\cite{angelani2000saddles,angelani2002quasisaddles,doye2003comment,broderix2000energy}, 
 thus connecting a landscape property to the dynamical MCT transition temperature.

 The actual divergence of the dynamical properties occur at a much lower temperature.
 At low temperatures, the dynamical properties like relaxation time ($\tau$) or viscosity ($\eta$)
 grow rapidly with decreasing temperature and can be fitted to a well known Vogel-Fulcher-Tammann (VFT) form which may be written as,
\begin{equation}
\tau(T)=\tau_{o}\exp \left[ \frac{1}{K_{VFT}(\frac{T}{T_{VFT}}-1)}\right],
\label{VFT}
\end{equation}
\noindent
where $K_{VFT}$ is the kinetic marker for the growth rate known as fragility, and $T_{VFT}$ is the temperature where the relaxation time diverges.
The laboratory glass transition temperature $T_g$ is higher than $T_{VFT}$, so $T_{VFT}$ is physically unattainable. Empirically 
it has been found that for a wide range of systems $T_{VFT}$ is close to the Kauzmann temperature $T_K$ where the configurational entropy goes to zero.
 The kinetic fragility ($K_{VFT}$) has also been connected to a thermodynamic quantity via Adam-Gibbs relationship 
\cite{adam-gibbs,sciortino-kob-tartaglia-prl,Srikanth_nature,speedy-1,speedy-2,foffi,stanley-nature,starr-sciortino-et-al,saika-voivod,mossa-tartaglia-pre,starr_douglas_sastry,shila-jcp}.
Thus it appears that all the dynamical temperatures have a thermodynamic connection.

  Apart from the ones described above, there are a few other connections between the dynamical transition temperatures and thermodynamics.
In a recent work,  we have reported that the configurational entropy can be divided into two parts - pair configurational entropy, $S_{c2}$, and residual multiparticle
entropy, $\Delta S$ \cite{bssb}. Interestingly the temperature at which $S_{c2}$ vanishes is found to be close to the MCT transition temperature \cite{unravel}.
%Through the well known Adam Gibbs (AG) relation we can connect dynamics to thermodynamics
The vanishing of pair configurational entropy ($S_{c2}$) implies that the pair 
relaxation times as obtained {\it via} the AG relation diverges at the MCT transition temperature. 
The AG relation is based on activated dynamics and MCT is a mean field theory which at the two body level does not address any activation. Also note that 
the microscopic MCT does not have any apparent connection to entropy. 
Despite the absence of any such connection, our study revealed that the vanishing of $S_{c2}$ is closely related to the MCT transition. 
In this way, we have further connected a dynamical transition temperature $T_c$ to a thermodynamic temperature $T_{K2}$.

In this present study we focus on the connection between the onset temperature and  thermodynamics.
As mentioned earlier, from the dynamic point of view,
this temperature is usually
associated with the non Arrhenius temperature dependence of the relaxation time and the stretched exponential behaviour of the time correlation function.
It is also identified from the onset of plateau behaviour of  self intermittent scattering function and  the Stokes Einstein breakdown temperature
\cite{sengupta2013breakdown,hocky2014crossovers,flenner2014universal,szamel-ntw}.
From the thermodynamic point of view this is the temperature where inherent structure potential energy ($e_{IS}$) starts to drop\cite{srikanth}
and the nature of the local potential energy minima changes \cite{sastry2000onset}.
Sciortino {\it et al.} \cite{sciortino-kob-tartaglia-prl,sciortino2000thermodynamics} have found that in the supercooled state, the inherent structures almost completely decouples from ``vibrational'' thermodynamics and have identified
a temperature as onset below which this phenomenon occurs.

  Brumer et al\cite{Reichman}, while connecting the onset to the microscopic MCT transition temperature, have reported a different value of $T_{onset}$
as obtained from fitting the inherent structure energy values to two straight lines.
 Thus, we can conclude that there exists an ambiguity in predicting the onset temperature using different protocols
given that they attempt to assign a particular temperature to the behaviour that may be characterized as a broad crossover. 
We find this onset temperature can be solely described from another thermodynamic quantity, the entropy. We observe that the temperature at which the two body entropy and
excess entropy undergo a crossing is related to the onset of  dynamical cooperativity
 in supercooled liquids. The residual multi particle entropy (RMPE)\cite{giaquinta-1,giaquinta-2,giaquinta-3,giaquinta-4}, the difference between the two body entropy 
and excess entropy, becomes positive below the crossing
temperature. In the present work we have studied a wide range of systems with varying densities, fragilities  to prove that our observation is generic in nature.

 The paper is organized as follows:
The simulation details are given in Sec. II. In Sec. III we describe
the methods used for evaluating the various quantities of interest and provide other necessary background. In Sec. IV we present the results and discussions.  
Sec. V contains a brief conclusion.

\section{Simulation Details}
In this study, we perform extensive molecular dynamics simulations for three-dimensional binary mixtures in 
the canonical ensemble. The system contains $N_A$ particles of type A and $N_B$ particles of type B under periodic boundary conditions. The total number 
density is fixed at $\rho=N/V$ with the total number of particles $N = N_A + N_B$ and a system volume $V$.
The models studied here, are the well-known models of glass-forming liquids: the binary Kob-Andersen Lennard-Jones (LJ) liquids \cite{kob}and  the 
corresponding WCA version (WCA) \cite{chandler}, the binary Wahnstr\"{o}m (WAHN) liquids \cite{wahnstrom} and a network-forming (NTW) \cite{coslovich_pastore_jpcm_ntw}
liquid that mimics
$SiO_2$ with short-range spherical potentials. The molecular dynamics (MD) simulations have been carried out using the LAMMPS 
package \cite{lammps}. For all state points, three to five independent samples with run lengths $>$ 100$\tau$ ($\tau$ is the $\alpha$-
relaxation time) are analyzed.

\subsection{ KALJ and KAWCA: binary mixture of Kob Andersen Lennard-Jones particles and corresponding WCA version}

 The most frequently studied model for glass forming liquids is Kob-Andersen model which is a binary mixture (80:20) of Lennard-Jones (LJ) particles \cite{kob}.
 The interatomic pair  
potential between species $\alpha$ and $\beta$, with ${ \alpha,\beta}= A,B$, 
$U_{\alpha\beta}(r)$ is described by a shifted and truncated Lennard-Jones (LJ) potential, as given by:
\begin{equation}
 U_{\alpha\beta}(r)=
\begin{cases}
 U_{\alpha\beta}^{(LJ)}(r;\sigma_{\alpha\beta},\epsilon_{\alpha\beta})- U_{\alpha\beta}^{(LJ)}(r^{(c)}_{\alpha\beta};\sigma_{\alpha\beta},\epsilon_{\alpha\beta}),    & r\leq r^{(c)}_{\alpha\beta}\\
   0,                                                                                       & r> r^{(c)}_{\alpha\beta}
\end{cases}
\label{LJ_pot}
\end{equation}

\noindent where $U_{\alpha\beta}^{(LJ)}(r;\sigma_{\alpha\beta},\epsilon_{\alpha\beta})=4\epsilon_{\alpha\beta}[({\sigma_{\alpha\beta}}/{r})^{12}-({\sigma_{\alpha\beta}}/{r})^{6}]$ and
 $r^{(c)}_{\alpha\beta}=2.5\sigma_{\alpha\beta}$ for the LJ systems (KALJ) and $r^{(c)}_{\alpha\beta}$  is equal to the position of the minimum of $U_{\alpha\beta}^{(LJ)}$
for the WCA systems (KAWCA) \cite{chandler}. Length, temperature and
time are given in units of $\sigma_{AA}$, ${k_{B}T}/{\epsilon_{AA}}$ and $\tau = \surd({m_A\sigma_{AA}^2}/{\epsilon_{AA}})$, 
respectively.  Here we have simulated Kob Andersen Model  
with the interaction parameters  $\sigma_{AA}$ = 1.0, $\sigma_{AB}$ =0.8 ,$\sigma_{BB}$ =0.88,  $\epsilon_{AA}$ =1, $\epsilon_{AB}$ =1.5,
 $\epsilon_{BB}$ =0.5, $m_{A}$ = $m_B$=1.0. We have performed MD simulations in the canonical ensemble (NVT) using  Nos\'{e}-Hoover thermostat  with integration timestep 0.005$\tau$.
 The time
constants for  Nos\'{e}-Hoover thermostat  are taken to be 100  timesteps.
The sample is kept in a cubic box with periodic boundary condition.
 System size is $N = 500$, $N_A = 400$ (N $=$ total number
of particles, $N_A$ $=$ number of particles of type A) and we have studied a broad range of density $\rho$ from 1.2 to
1.6.
\subsection{WAHN: : binary mixture of Wahnstr\"{o}m Lennard-Jones particles}

 The Wahnstr\"{o}m model, which is a binary mixture (50:50) of Lennard-Jones (LJ) particles \cite{wahnstrom}, follows Lorentz Berthelot rule of mixing 
i.e. $\sigma_{AB}=(\sigma_{AA}+\sigma_{BB})/2 $
  and  $\epsilon_{AB}=\sqrt{{\epsilon_{AA}}\epsilon_{BB}}$. The interaction potentials are the 
same as in Eq. \ref{LJ_pot}, in which the size, mass, and energy
ratios are given as $\sigma_{AB}=\frac{11}{6}\sigma_{AA},\sigma_{BB}=\frac{5}{6}\sigma_{AA}
,  m_A=1,m_B =0.5 ,\epsilon_{AB}=\epsilon_{AA}=\epsilon_{BB}=1$. Length, temperature and
time are given in units of $\sigma_{AA}$, ${k_{B}T}/{\epsilon_{AA}}$ and $\tau = \surd({m_A\sigma_{AA}^2}/{\epsilon_{AA}})$, 
respectively. System size is $N = 500$, $N_A = 250$ (N $=$ total number
of particles, $N_A$ $=$ number of particles of type A). The number density of WAHN mixtures is
$\rho = 1.2959$.

\subsection{NTW: tetrahedral network-forming liquids}

We study a model of network-forming liquids \cite{coslovich_pastore_jpcm_ntw} interacting via spherical short-ranged potentials.
This model is simple model which mimics $SiO_2$ glasses, in which tetrahedral networks  dominate the dynamics.
The interaction potentials are given as

\begin{equation}
 U_{\alpha\beta}(r)=\epsilon_{\alpha\beta}[(\frac{\sigma_{\alpha\beta}}{r})^{12}-(1-\delta_{\alpha\beta})(\frac{\sigma_{\alpha\beta}}{r})^6].
\end{equation}
Here $\delta_{\alpha\beta}$ is the Kronecker delta function. The interaction is truncated and shifted at $r=2.5\sigma_{\alpha\beta}$. The size, mass, and energy
ratios are given as  $\sigma_{AB}/\sigma_{AA}=0.49, \sigma_{BB}/\sigma_{AA}=0.85, m_B/m_A=0.57, \epsilon_{AB}/\epsilon_{AA}=24
, \epsilon_{BB}/\epsilon_{AA}=1$.

Following the earlier work \cite{coslovich_pastore_jpcm_ntw}, we simulated a system where $N = 500$, $N_A = 165$ (N $=$ total number
of particles, $N_A$ $=$ number of particles of type A). The number density of NTW mixtures is
$\rho = 1.655$.

\section{Definitions and Background}
\subsection{Relaxation time}

We have calculated the relaxation times from the decay of the
overlap function q(t), from the condition $q(t)=
1/e$.  $q(t)$ is defined as

\begin{eqnarray}
  q(t) \approx \frac{1}{N}\left \langle \sum_{i=1}^{N} \delta({\bf{r}}_i(t_0)-{\bf{r}}_i(t+t_0)) \right \rangle.
\end{eqnarray}

Again, the $\delta$ function is approximated by a Heaviside step function $\Theta(x)$ which defines the
condition of “overlap” between two particle positions
separated by a time interval t:
\begin{eqnarray}
 q(t) \approx \frac{1}{N}\left \langle \sum_{i=1}^{N} \Theta (\mid{\bf{r}}_i(t_0)-{\bf{r}}_i(t+t_0)\mid) \right \rangle \nonumber\\
\Theta(x) = 1, x \leq {\text{a implying “overlap”}} \nonumber\\
=0, \text{otherwise}.
\end{eqnarray}

For the calculation of $q(t)$ the cut off parameter `a' is chosen as 0.3 for the KA model with LJ and WCA potentials \cite{shila-jcp} and for
the WAHN model. 
For NTW model, the cut off parameter for overlap function is taken as $a=0.2$ \cite{szamel-ntw}.

 Since relaxation times from $q(t)$ and $F_s (k, t)$
behave very similarly, we have used the time scale
obtained from q(t).

\subsection{Diffusivity}
Diffusivities (D ) are obtained from the mean squared
displacement (MSD) of the particles. We have calculated MSD as follows,
\begin{equation}
 R^2(t)=\frac{1}{N} \sum_{i} \langle({\bf{r}}_i(t)-{\bf{r}}_i(0))^2 \rangle, 
\end{equation}
 and from the long time behavior of MSD, the diffusion coefficient D can be written as,
\begin{equation}
 D= \lim_{t\rightarrow \infty} \frac{R^2(t)}{6t}.
\end{equation}
At longer time  by fitting the MSD with time,  we obtain D from the slope of the fitted plot.

 \subsection{Excess Entropy}

 The thermodynamic excess entropy, $S_{ex}$, is defined as the difference between
the total entropy $(S_{total})$ and the ideal gas entropy $(S_{id})$ at
the same temperature $(T)$ and density $(\rho$)for all the model
glass formers. The $S_{ex}$ has been evaluated using the procedure given in
Ref- \cite{Murari_charu_jcp}. The entropy was initially evaluated at a high temperature
and low reduced density for each model,
where the system can be assumed to behave as an ideal gas.
Entropies at any other state points, relative to this reference ideal
state point, can be calculated using a combination of isochoric
and isothermal paths, ensuring that no phase transitions occur
along the path. The entropy change along an isothermal path
is given by,
\begin{equation}
 S(T,V^\prime)- S(T,V)=\frac{U(T,V^\prime)- U(T,V)}{T}+\int_{V}^{V^\prime}\frac{P(V)}{T}dV,
\end{equation}
and along the isochoric path it is given by,

\begin{equation}
 S(T,V^\prime)- S(T,V)=\int_{T}^{T^\prime}\frac{1}{T} \left(\frac{\partial U}{\partial T}\right)_V dT.
\end{equation}

\subsection{Pair entropy}

As discussed earlier the excess entropy $S_{ex}$, defined as the difference between the total entropy ($S_{total}$) and the ideal gas entropy ($S_{id}$ ) at
the same temperature ($T$) and density ($\rho$), can also be expanded in an infinite series, 
$S_{ex}=S_{2}+S_{3}+.....=S_{2}+\Delta S$ using Kirkwood factorization \cite{Kirkwood} of the N-particle distribution function \cite{green_jcp,raveche,Wallace}. 
 $S_{n}$ is the $``n"$ body contribution to the entropy. Thus the pair excess entropy is
  $S_{2}$ and the higher order contributions to excess entropy is given by 
the residual multiparticle entropy (RMPE), $\Delta S=S_{ex}-S_{2}$  \cite{giaquinta-1,giaquinta-2,giaquinta-3,giaquinta-4}.
The pair entropy $S_{2}$ for a binary system can be written in terms of 
the partial radial distribution functions,

\begin{widetext}
\begin{equation}
\frac{S_{2}}{k_{B}}=-2\pi\rho \sum_{\alpha,\beta}x_{\alpha} x_{\beta} \int_0^{\infty} \{g_{{\alpha}{ \beta}}(r) \ln g_{{\alpha} {\beta}}(r)- [g_{{\alpha}{\beta}}(r)-1]\} r^2 d {r},
\label{s2_final}
\end{equation}
\end{widetext}
\noindent  where $ g_{{\alpha}{ \beta}}(r)$ is the atom-atom pair correlation between atoms of type $\alpha$ and $\beta$, $\rho$ is the density of the system,
 $x_{\alpha}$ is the mole fraction of component $\alpha$ in the mixture, and $k_B$ is the Boltzmann constant.

\section{Results}

In this paper we mainly focus on the determination of the onset temperature from the information of different components of the entropy. However, before we do that we review
the existing methods of $T_{onset}$ determination.
\subsection{Earlier methods: From Dynamics}
 The onset temperature has been identified as a temperature where
 the temperature dependence of relaxation time changes from  Arrhenius to super-Arrhenius behaviour \cite{srikanth}.
At high temperatures, the time dependence of $F_s(k,t)$ is exponential, whereas at lower temperatures, it can be fitted to the Kohlrausch-Williams-Watts stretched
exponential form given by,
\begin{equation}
 F_s(k,t)=\exp \left[ (t/\tau(T))^{\beta(T)}\right],
\end{equation}
 where the exponent, $\beta < 1$ \cite{srikanth,sastry2000onset,sastry1999potential}. 
For KALJ model at $\rho =1.2$ it has been identified as $T_{onset}=1.0$.

\begin{figure}[h]
\centering
\includegraphics[width=0.45\textwidth]{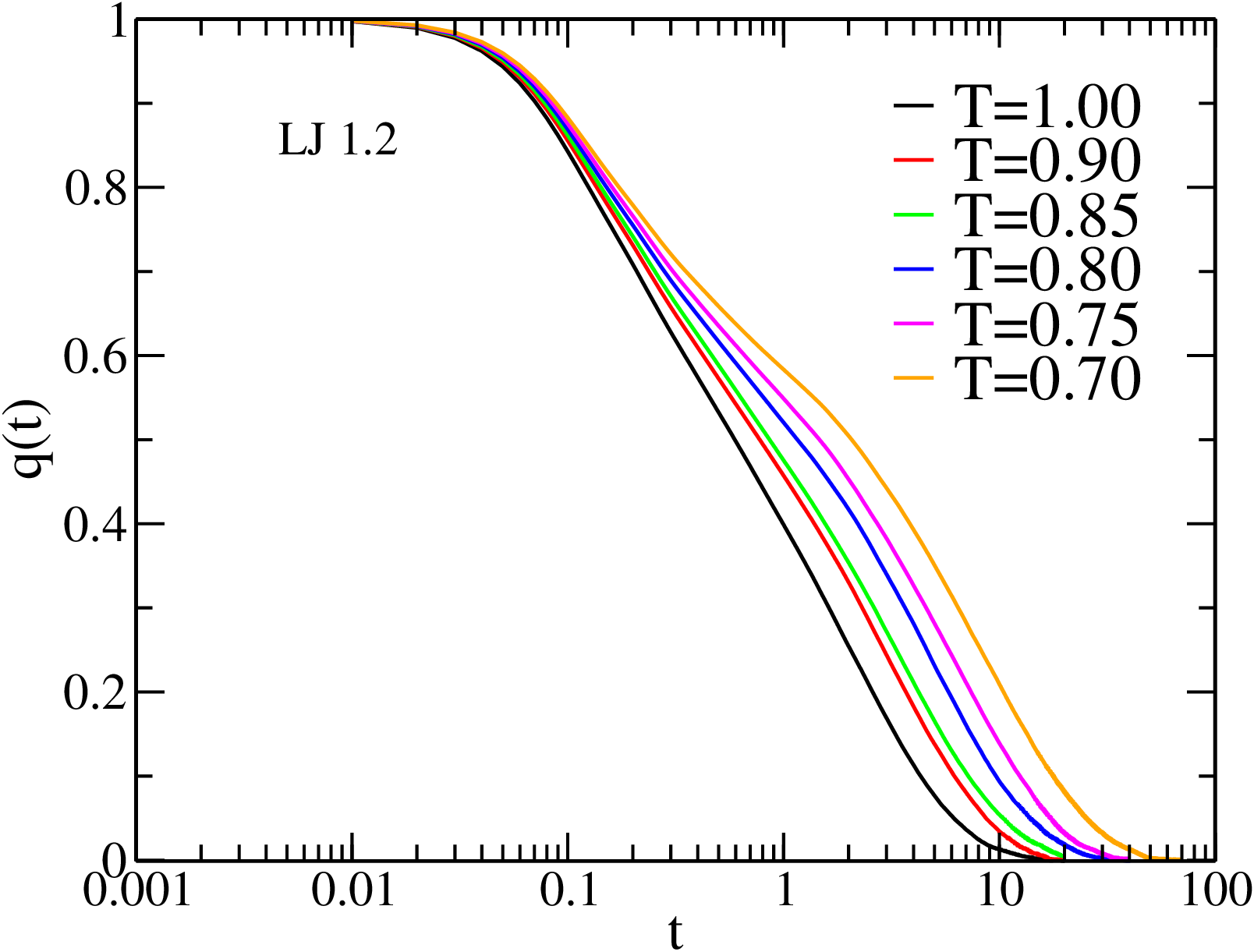}
\caption{ Overlap function $q(t)$ at different temperatures for LJ system at $\rho=1.2$. From this figure, it is difficult to identify a particular temperature at which the plateau 
arises.}
\label{onset-fskt}
\end{figure}

 Kob {\it et al.} \cite{kob-2} have identified the onset temperature from the first time appearance of the plateau in the time correlation function. They have reported that  for the KA model at 
$\rho=1.2 $ the plateau appears at $T=0.8$. 
Chakravarty {\it et al.} have identified the $T_{onset}$ from the onset of the plateau in MSD and for KA model at the same density this has a higher value \cite{Murari_charu_jcp}.
We plot $q(t)$ of  KA model at $\rho=1.2$
for different temperatures in Fig. \ref{onset-fskt}. As shown in Fig. \ref{onset-fskt}, it is difficult to exactly identify a particular temperature at which the plateau begins to appear.  
The value of $T_{onset}$ from the appearance of the plateau for the different systems are given in Table\ref{crossing-temperature-table}.

\subsection{Earlier Method: From potential energy landscape}

\begin{figure}[h]
\centering
%\subfigure{
{
 \includegraphics[width=0.45\textwidth]{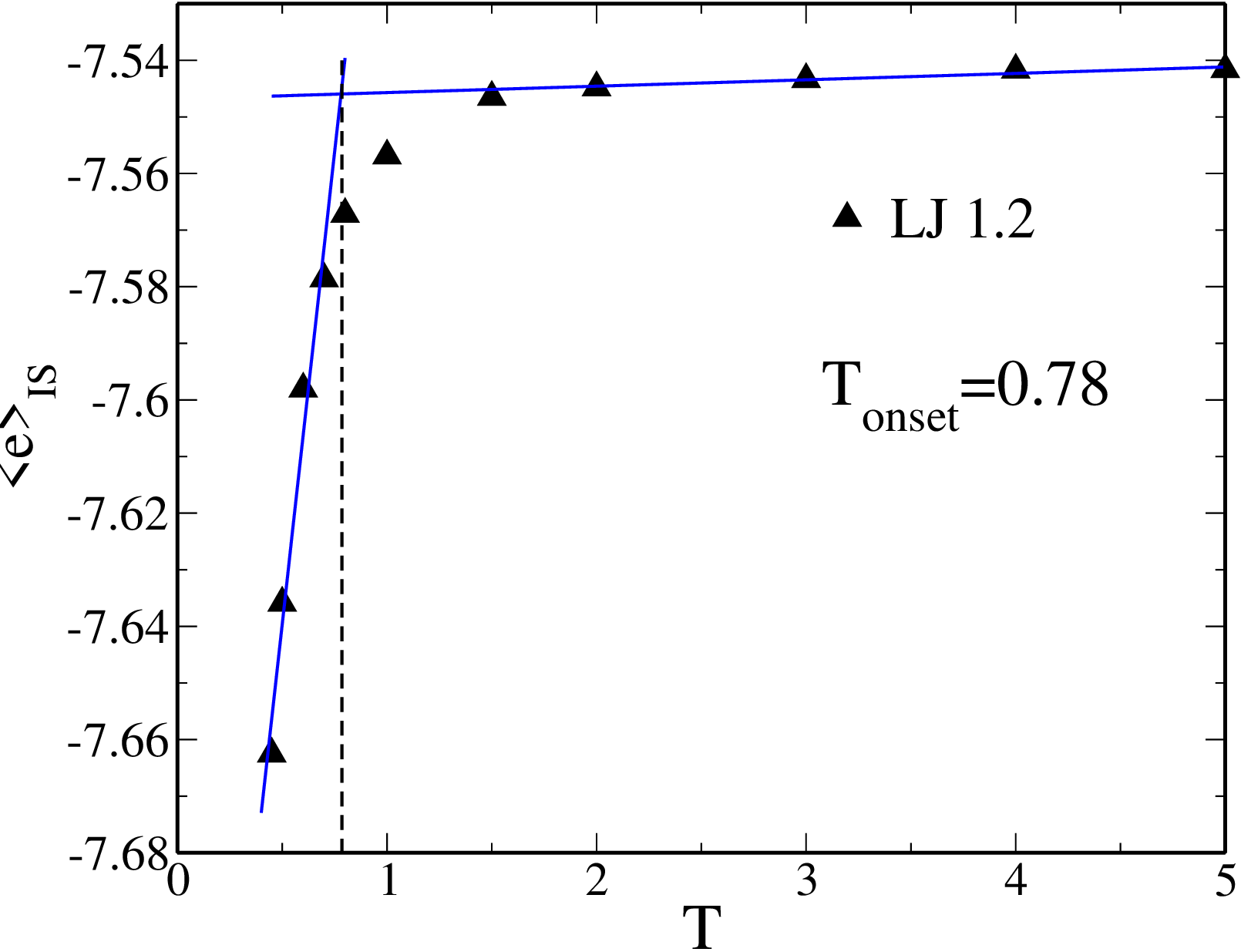}}
\caption{\it{Average inherent structure energy ($<e_{IS}>$) vs. temperature for KA LJ model. The solid lines are straight line fits to high and low temperature slopes.
The onset temperature can be defined from the point where the two lines undergo crossover \cite{Reichman} or from the drop of average inherent structure energy
\cite{srikanth}. Following the same protocol as mentioned in Ref-\cite{Reichman} we find $T_{onset}=0.78$ for KA model at $\rho=1.2$. Note that, the onset
temperature determination from this method is extremely fitting depended.  }}
\label{ns-fit}
\end{figure}

As mentioned earlier the onset temperature has also been connected to energy landscape\cite{srikanth,sastry2000onset}.
Sastry {\it et al.} \cite{srikanth} demonstrated that below the onset temperature, the average inherent structure potential energy ($<e_{IS}>$) of the systems decreases drastically 
with decreasing temperature, and above it, the $<e_{IS}>$ is a constant. They have argued that above $T_{onset}$ the dynamics is  landscape independent
 and below it, the dynamics is 
``landscape influenced''.  For KA LJ model at $\rho=1.2$, they have identified $T_{onset}=1.0$ \cite{srikanth}.
It has also been shown that there is a change in the nature of local potential energy minima at this temperature \cite{sastry2000onset}.
Brumer and Reichman have followed the same protocol and identified the onset temperature for KA model at $T=0.9$ by fitting two straight lines
 to high and low temperature regions \cite{Reichman}. 
 They have also solved the microscopic MCT using the inputs of the simulated static correlation function and have shown that the transition temperature
as predicted by this theory is higher than that predicted by  fitting  the simulated dynamics to the MCT predicted power law behaviour\cite{Reichman}. Moreover,
they have shown that the transition temperature predicted by microscopic MCT is similar to the onset temperature.
 However, note that the identification of the onset temperature is fitting dependent. As shown in Fig.\ref{ns-fit} we perform a similar fitting exercise, and by changing  the fitting range
we can identify the onset temperature near $0.78$.

Sciortino {\it et al.} \cite{sciortino2000thermodynamics} have shown that below a certain temperature, the harmonic basin approximation used to calculate
the vibrational entropy is valid. They have identified this temperature as onset temperature below which the 
 the thermodynamics of the inherent structures  decouples
from the ``vibrational'' thermodynamics. They have used the same basin entropy estimate to infer the configurational entropy and
 at high temperatures, a harmonic basin entropy estimate fails. 
Thus they have identified this $T=0.8$ as an onset temperature for KA model.

 Note that all the methods require the same inherent structure energy as input, but predicts different temperatures as ``onset''. Thus, 
in the literature even in the determination of $T_{onset}$ from thermodynamics, there exists an ambiguity.
The values of $T_{onset}$ for different systems as obtained by fitting $e_{IS}$ to two different straight lines are given in Table \ref{crossing-temperature-table}.

\begin{widetext}
 \begin{table}[h]
 \centering
\caption{The value of onset temperatures as obtained from different methods for different systems.}
\begin{tabular}{ l| r r| r r |r r|r|r  }
 \hline
\hline

& &$\rho=1.2$ & & $\rho=1.4$& &  $\rho=1.6$&WAHN&NTW \\
\hline
& LJ & WCA& LJ & WCA& LJ & WCA &&\\
 \hline
  $T_{Plateau-arises}$ &0.9-0.8  & 0.7-0.6& 1.7-1.6& 1.6-1.4& 3.25-2.75 &3.25-2.75&1.0-0.9& 0.51-0.44 \\
  $T_{e_{IS}-drop}$ & 0.9-0.78 & 0.7-0.6& 1.6-1.5& 1.5-1.4& 2.80-2.85&2.8-3.75&0.8-0.75& 0.55-0.5 \\
  $T^{SE}_{breakdown}$ &0.781  & 0.609& 1.57& 1.43& 2.782 &2.707&0.8& 0.492 \\
  $T_{crossing}$& 0.77  & 0.61&1.50  &  1.43& 2.86&2.85 &0.86&0.58\\
\hline
\hline
\end{tabular}
\label{crossing-temperature-table}
\end{table}
\end{widetext}

\subsection{Earlier Method: From Stokes Einstein Breakdown}

According to SE relation the $D\eta /T$ should be independent
of the temperature  where $D$ is the diffusion coefficient and $\eta$ is the viscosity.
Since via simulation studies it is demanding to calculate the $\eta$, often $\eta$ is replaced by the relaxation time, $\tau$.
In literature it has been found that $\eta \propto \tau/T$ \cite{chong2009coupling,mezei1987neutron,yamamoto1998heterogeneous}, hence  $D\tau $  is  
treated as a numerical quantity to investigate the validity of the SE relation. 
In the last few decades, several experiments and simulation studies have conclusively shown that the SE relation breaks down at low temperature
\cite {pollack1981atomic,chang1997heterogeneity,
swallen2011self,kob1994scaling,tarjus1995breakdown}.
Although some studies claim that the breakdown of the SE relation occurs around the mode coupling temperature $T_c$ \cite{ediger2000spatially},
a couple of simulation studies show that it is connected to the crossover from high-temperature to low-temperature behavior,
 referred to as the onset temperature of slow dynamics\cite{srikanth,sengupta2013breakdown}.

\begin{figure}[h]
\centering
\subfigure{
\includegraphics[width=0.45\textwidth]{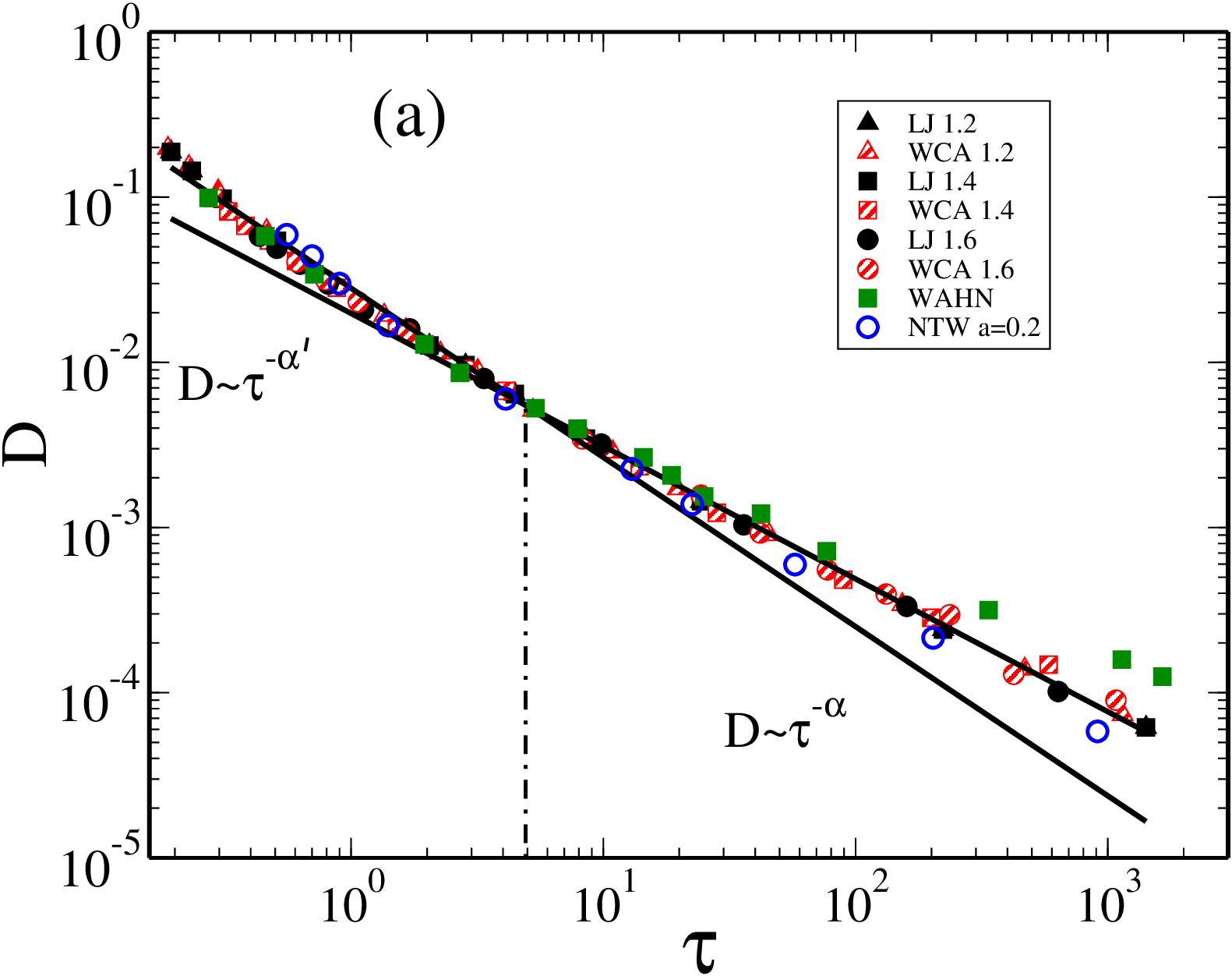}}
\subfigure{
\includegraphics[width=0.45\textwidth]{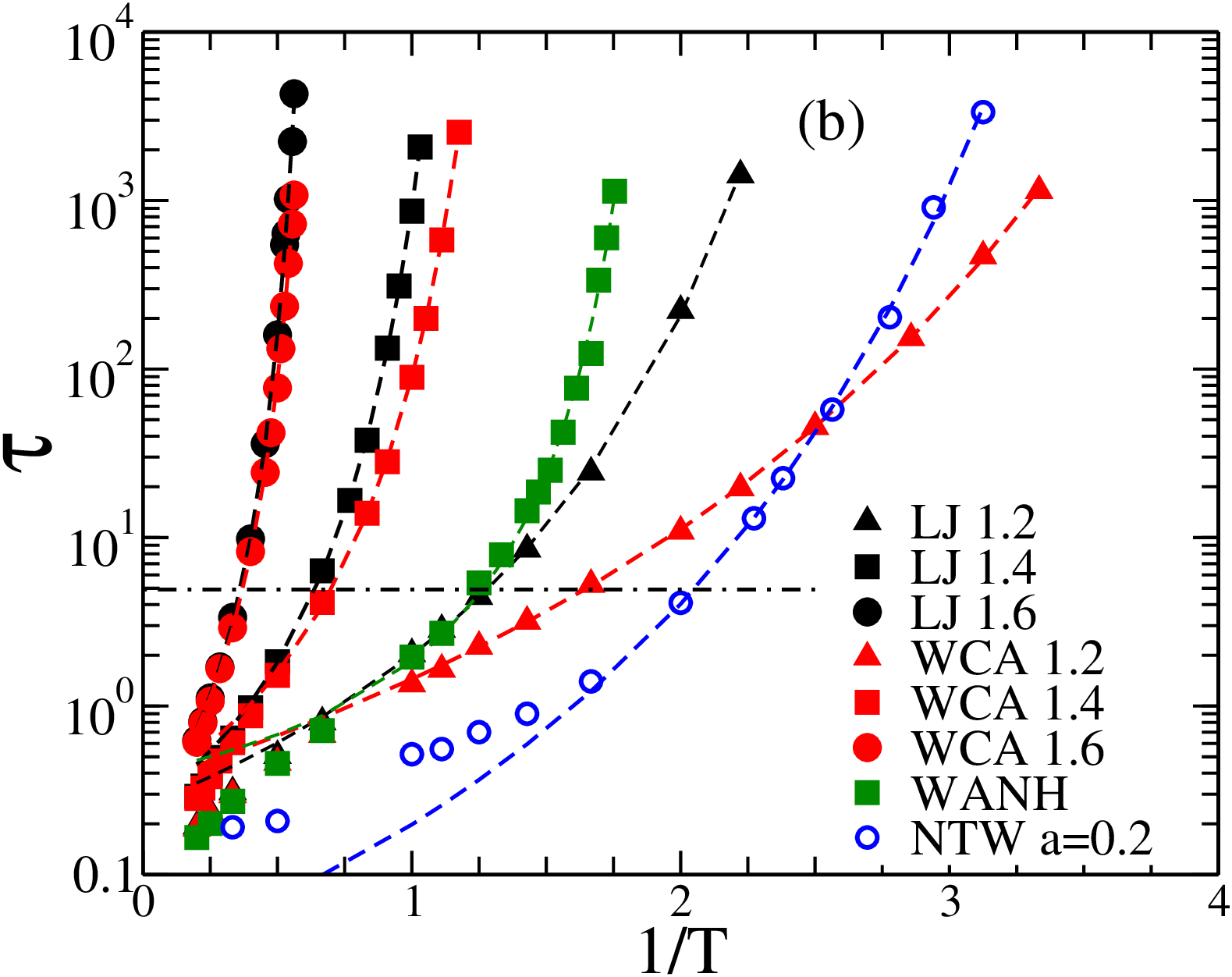}}
\caption{ (a)Diffusion coefficient vs relaxation time for all the systems. The solid lines are fit of the data to $D\approx\tau^{-\alpha}$ at high 
temperature and $D\approx\tau^{-\alpha\prime}$ at low temperature. The exponent values are given in Table \ref{alpha-table}. We find the SE breakdown occurs almost at a
particular value of relaxation time for all the systems. Note that the SE breakdown temperature is also fitting dependent.
 (b) The plot of relaxation time ($\tau$) vs. temperature for different systems. The dashed lines are VFT fits. We can identify the temperature at which the SE breakdown occurs. The vertical dash dotted 
line in (a) corresponds to the horizontal dash dotted line in (b).   
}
\label{se-breakdown}
\end{figure}

   In this present work, we study the SE breakdown for KA model with LJ and WCA potentials at different densities, WAHN model at $\rho=1.2959$ and NTW model at 
$\rho=1.655$. 
 We plot SE breakdown in Fig. \ref{se-breakdown}a and the fitting parameters are given in the Table \ref{alpha-table}.
At high temperature the systems follow $D \sim \tau^{-\alpha}$ where the exponent $\alpha\sim 1$ for all the systems, supporting the validity of  SE relation
 at high temperatures.
However at low temperatures the exponent $\alpha\prime$ in the relation $D\sim\tau^{-\alpha\prime}$  shows a deviation from unity. The degree of the departure of the
value of $\alpha\prime$ from unity gives 
a measure of the  breakdown. The value of $\alpha\prime$ and its connection to the fragility will be addressed later.
As reported earlier \cite{flenner2014universal}, the SE breakdown for all the systems happen at a particular value of alpha-relaxation time ($\tau\approx 4.92$ 
as seen in Fig.\ref{se-breakdown}a).
The breakdown temperature is identified from the temperature dependence of relaxation time as given in Fig.\ref{se-breakdown}b. 
 We have  tabulated the SE breakdown temperatures in Table \ref{crossing-temperature-table}. The breakdown temperature for the LJ system at $\rho=1.2$ is close to that reported
by Sengupta {\it et al.} \cite{sengupta2013breakdown}, but higher than that predicted by Szamel {\it et al.} \cite{flenner2014universal}. 
 Note that, this result is also completely fitting dependent. For NTW model the breakdown temperature obtained from our study is similar to that predicted 
by Szamel {\it et al.} \cite{szamel-ntw} by using the viscosity as a marker for solvent dynamics.
 Note that in a recent study it has been shown that the SE breakdown temperature depends on the wave number at which the relaxation time is calculated.
The breakdown temperature decreases with the wave number \cite{Anshul-k-dependence} However, the temperature, at which the SE relationship breaks down between
diffusion coefficient and viscosity, is similar to the onset temperature.

Note that our study of SE breakdown also shows that the degree of SE breakdown is correlated with fragility. We find that the systems which are more fragile show 
a larger departure from SE prediction (note the values of $\alpha^\prime$ in Table \ref{alpha-table}). 
 This is similar to that found for the square well fluids \cite{Anshul-jpcb}.
For the NTW model the difference between $\alpha$ and $\alpha\prime$ is very little. 
This is similar to the observation by Mishra {\it et al.} \cite{mishra2013two} where they have
found that for strong liquids there is no change in exponent in the SE relation.

 \begin{table}[h]
 \centering
\caption{The exponents obtained from Fig. \ref{se-breakdown} have been given below. The breakdown is maximum ($\alpha^\prime$ is lowest )for the WAHN system which is most fragile among the systems
studied here and the breakdown is minimum ($\alpha^\prime$ is highest) for NTW model which the strongest system among all. }
\begin{tabular}{ l| r r| r r |r r  |r|r }
 \hline
\hline

& &$\rho=1.2$ & & $\rho=1.4$& &  $\rho=1.6$& WAHN & NTW\\
\hline
& LJ & WCA& LJ & WCA& LJ & WCA &- &a=0.2\\
 \hline
  $\alpha$ &1.024  & 1.061& 1.032& 0.987& 0.981 &0.987&1.001&1.030\\
 $\alpha^\prime$ &0.803 & 0.793& 0.788 &0.777 &0.808 &0.772&0.6139&0.851\\
\hline
\hline
\end{tabular}
\label{alpha-table}
\end{table}

\begin{widetext}
\begin{figure}[ht!]
\centering
\subfigure{
 \includegraphics[width=0.3\textwidth]{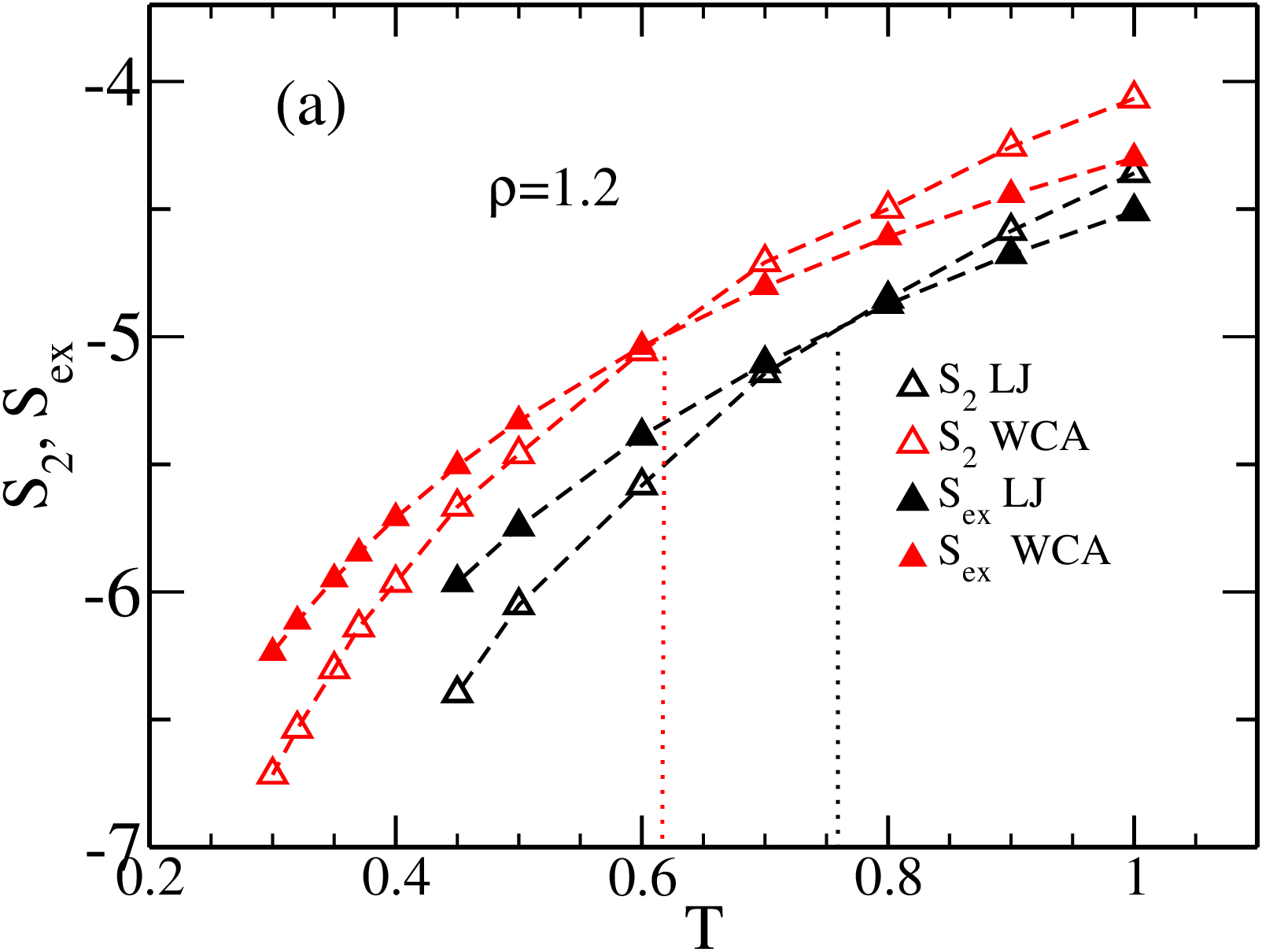}}
\subfigure{
 \includegraphics[width=0.3\textwidth]{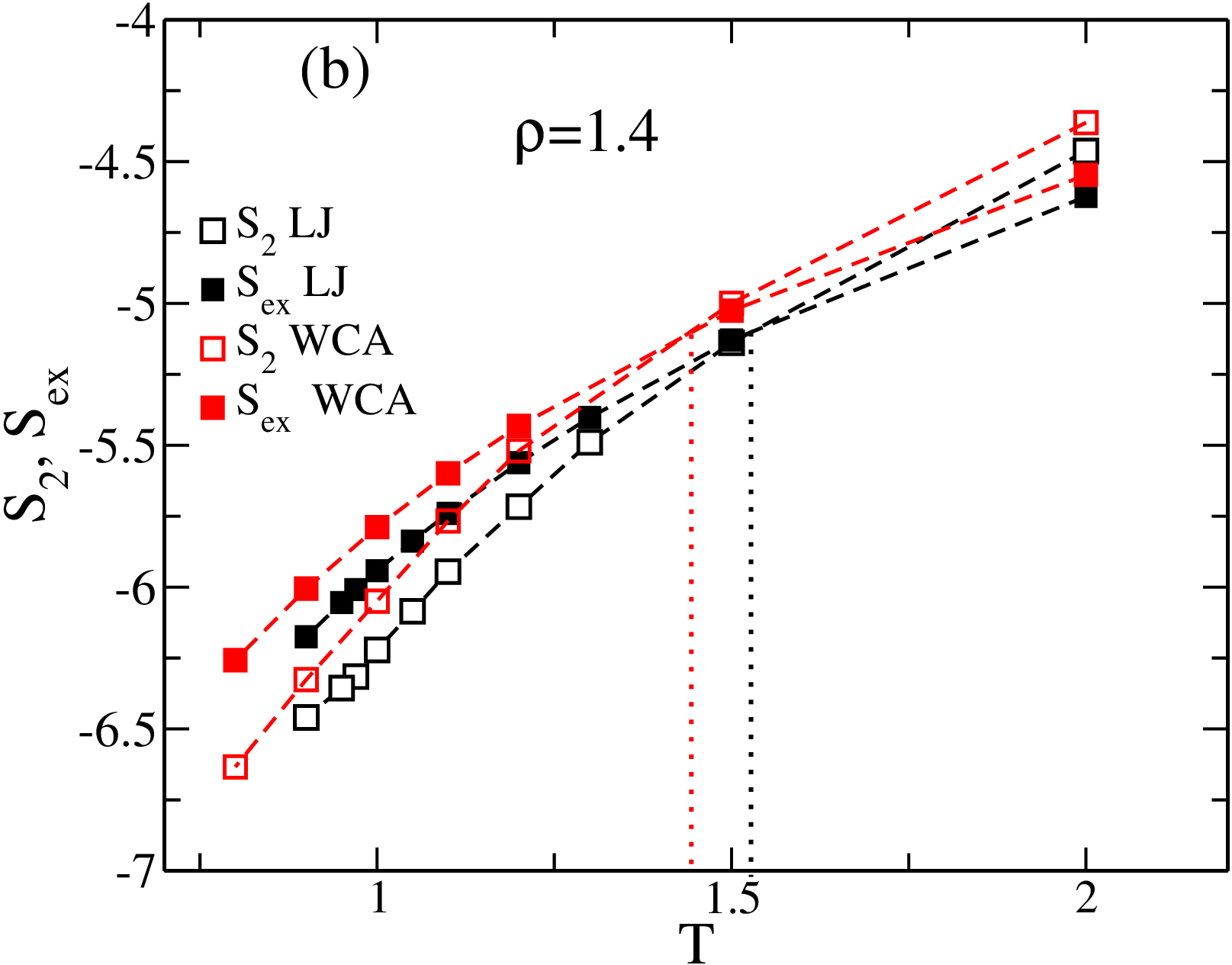}}
\subfigure{
 \includegraphics[width=0.3\textwidth]{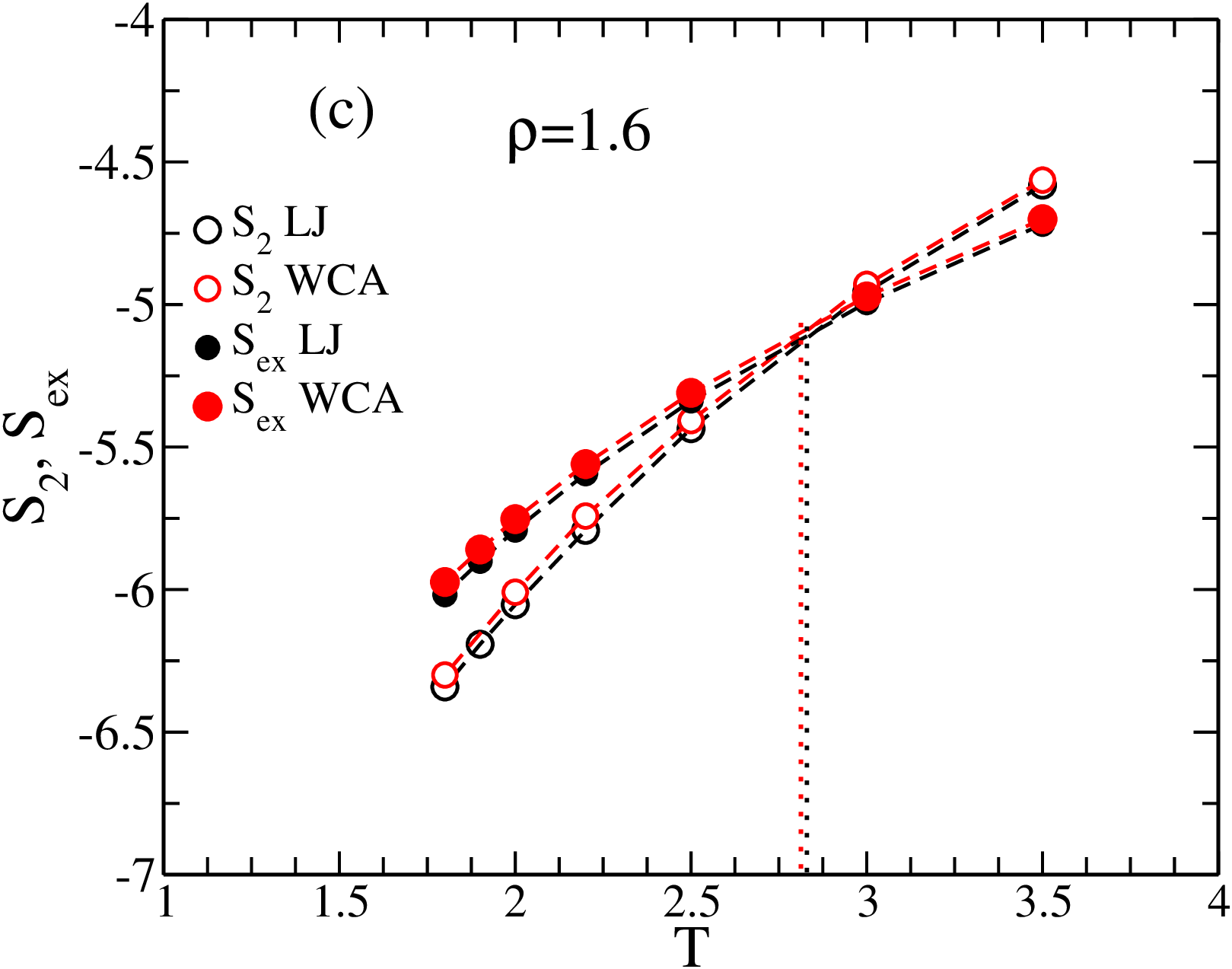}}
\subfigure{
 \includegraphics[width=0.3\textwidth]{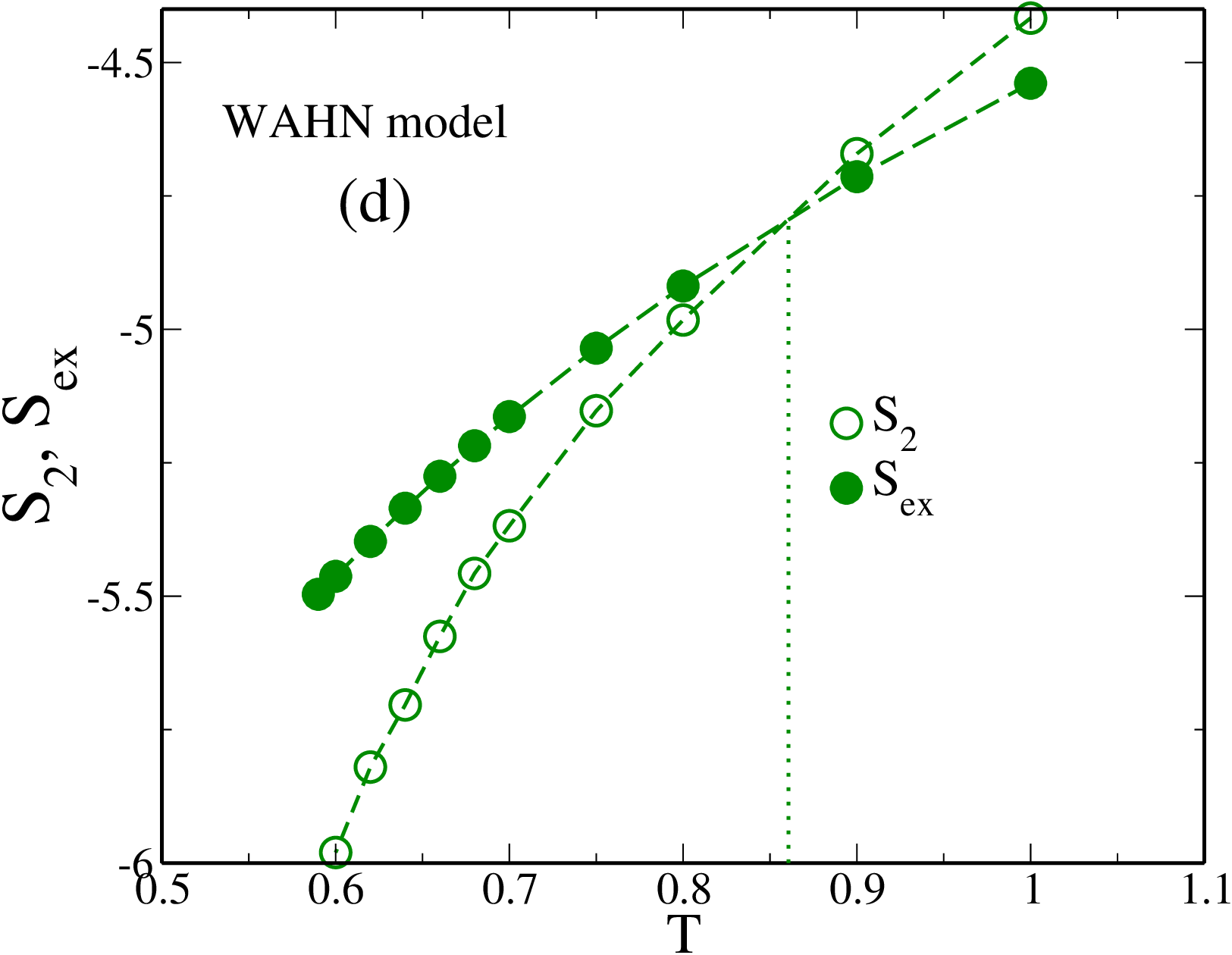}}
\subfigure{
 \includegraphics[width=0.3\textwidth]{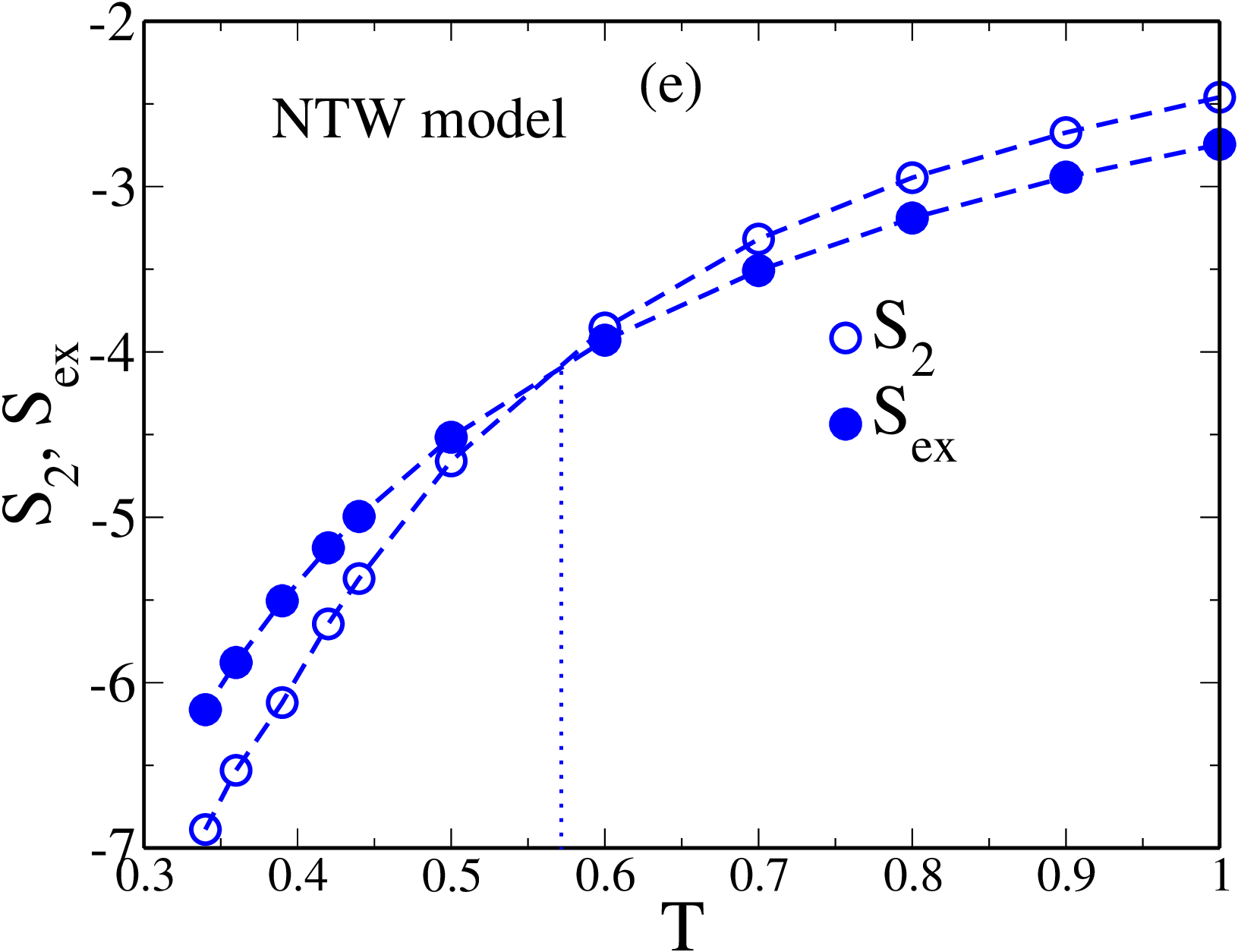}}
\caption{\it{The crossing between $S_2$ and $S_{ex}$ for different models- (a-c) KA model with LJ and WCA potentials at different densities, (d) WAHN model, (e) NTW model.
As density increases, the values of crossing temperature for KALJ and KAWCA systems
come closer. The crossing temperature can be identified as the ``onset temperature''.}}
\label{crossover_plot}
\end{figure}
\end{widetext}

\subsection{Present Method: Connection with Entropy}
Here we aim to explore a connection between onset temperature and entropy. The present method appears to be less ambiguous and is independent of numerical fittings.
We have calculated the excess entropy and pair entropy as mentioned in the earlier section (Sec-III).
As we lower the temperature, we find that both the entropies decrease at different rates. 
The common belief is that the many body interaction slows down the dynamics, thus $S_{2}$ should be higher than $S_{ex}$. 
As expected, at high temperatures $S_2$ is higher than $S_{ex}$, however 
as temperature decreases the $S_2$ becomes smaller than $S_{ex}$. This gives rise to a crossing between the two entropies. Thus the RMPE, $\Delta S$,  undergoes a change 
in sign, being negative at high temperatures and positive at low temperatures. We identify the temperature at which  $\Delta S =0$, where $S_2$ and $S_{ex}$ 
have a crossing, as the ``onset temperature''.
Note that the change in sign of $\Delta S$ predicts a role reversal of multiparticle entropy in the dynamics \cite{bssb}. 
The positive value of RMPE  means that it contributes to the increase in the entropy. Thus,  
the effect of RMPE on dynamics is similar to activation
and helps the system to explore more phase space.  Note that the temperature corresponding to zero RMPE has also been identified as the freezing temperature
in three dimension \cite{giaquinta-1,giaquinta-2,giaquinta-4,trusket_2006}.
            In Fig.\ref{crossover_plot} we plot $S_2$ and $S_{ex}$ for all the models and tabulate the crossing temperature in Table \ref{crossing-temperature-table}. Within the error
bar of $S_2$ and $S_{ex}$ calculation we can precisely identify the $T_{crossing}$.
 We like to emphasis that this method is independent of any fitting procedure.
We find that as density increases the crossing temperatures of the LJ and WCA systems come closer
which is consistent with our earlier observation where we find that different
dynamical and thermodynamical quantities for LJ and WCA systems come closer as the density increases \cite{post_prl_long}. For all the systems, we can identify
this crossover temperature as the temperature where the plateau in $q(t)$ first appears (not shown here) and the SE relation shows a breakdown.
Our result is similar to an earlier observation where it has been shown that the growth of domains of locally preferred structures (which should lead to many body
correlation) signals the onset of slow-dynamics \cite {coslovich}.

\section{Conclusion}

 While cooling, the behaviour of a liquid undergoes a transition at a temperature below which the landscape properties of the system starts
to contribute to its dynamics\cite{srikanth}. This is often referred to as the onset temperature. The identification of the onset temperature is tricky and there have been multiple 
routes to identify the temperature both using thermodynamical and dynamical properties of the system. In this work we review the existing methods of the onset temperature determination 
and find that there exists an ambiguity in determining it both from  dynamics as well as from thermodynamics. In most of the cases, the methods require  fitting
parameters which reduce the accuracy.

   In this present study we explore a connection between the onset temperature and the entropy for a wide range of systems by varying density and fragility.
We independently calculate the excess entropy, $S_{ex}$ and its pair counterpart, $S_2$. The difference between these two entropies is considered as the residual multiparticle entropy
(RMPE), which has contribution from higher order correlations. At high temperature, since the higher order correlations slow down the dynamics, the $S_{ex}$ is smaller
than the $S_2$. However, the temperature dependence of $S_{ex}$ and $S_2$ are different. At a certain temperature, they cross each other and the $S_{ex}$ becomes 
higher than $S_2$. Thus, the RMPE undergoes a change in sign, being negative at high temperature and positive at low temperature.
The role of RMPE in the dynamics also changes with the change in the sign. At high temperature where RMPE is negative, it slows down the dynamics.
The positive value of RMPE actually speeds up the dynamics over the value predicted by the two body part \cite{bssb}. Thus, its effect on the dynamics is similar 
to that of activation. Positive value of RMPE increases the entropy of the system and activation helps the system to overcome  the barrier in the landscape and eventually 
explore more phase space. Note that activation is also believed to be a many body process.
We identify the temperature of zero residual multiparticle entropy as the ``onset temperature'' below which the RMPE is positive. 
In this present method  we do not use any fitting parameter to identify the onset, thus it is expected to be a less ambiguous method.

    Note that in 3-D the $\Delta S=0$ is also connected to the freezing temperature\cite{giaquinta-1}. However it has been shown that this connection does not exist in other dimensions
\cite{trusket_2006}. It will
be interesting to study the correlation between $T_{onset}$ and $T_{crossing}$ in other dimensions.

 As mentioned
above we expect the RMPE to play the role of activation in the dynamics. It will also be interesting to understand the role of RMPE in systems which are mean field in nature
and where activation is known not to play a dominant role.

\clearpage
\end{document}